\documentclass[twocolumn,english,showpacs,preprintnumbers,amsmath,amssymb,floatfix,nofootinbib]{revtex4-1}
\usepackage[T1]{fontenc}
\usepackage[latin9]{inputenc}
\usepackage{color}
\usepackage{array}
\usepackage{amstext}
\usepackage{graphicx}
\usepackage{esint}
\usepackage{rotating}
\usepackage{slashed}
\usepackage{siunitx}
\usepackage[font=small,labelfont=bf]{caption}

\usepackage{epsfig}
\usepackage{graphics}

\usepackage{xcolor}
\usepackage[colorlinks = true,
linkcolor = magenta,
urlcolor  = blue,
citecolor = red,
anchorcolor = blue]{hyperref}
\makeatletter

\@ifundefined{textcolor}{}
{%
 \definecolor{BLACK}{gray}{0}
 \definecolor{WHITE}{gray}{1}
 \definecolor{RED}{rgb}{1,0,0}
 \definecolor{GREEN}{rgb}{0,1,0}
 \definecolor{BLUE}{rgb}{0,0,1}
 \definecolor{CYAN}{cmyk}{1,0,0,0}
 \definecolor{MAGENTA}{cmyk}{0,1,0,0}
 \definecolor{YELLOW}{cmyk}{0,0,1,0}
 }

\@ifundefined{definecolor}
 {\usepackage{color}}{}
\@ifundefined{definecolor}
 {\usepackage{color}}{}
\makeatother
\usepackage{babel}

\def\Re{{\cal R \mskip-4mu \lower.1ex \hbox{\it e}\,}}
\def\Im{{\cal I \mskip-5mu \lower.1ex \hbox{\it m}\,}}

\def\tev{\,{\ifmmode\mathrm {TeV}\else TeV\fi}}
\def\gev{\,{\ifmmode\mathrm {GeV}\else GeV\fi}}
\def\mev{\,{\ifmmode\mathrm {MeV}\else MeV\fi}}

\begin{document}

%%%%%%%%%%%%%%%%%%%%%%%%%%%%%%

\title{Transverse momentum dependent of charged pion, kaon and proton/antiproton fragmentation functions from $e^+e^-$ annihilation process }
	
%{\color{blue}Belle data}}
%in $e^+e^-$ annihilation process }

\author{Maryam Soleymaninia$^{1}$}
\email{Maryam\_Soleymaninia@ipm.ir}

\author{Hamzeh Khanpour$^{2,1}$}
\email{Hamzeh.Khanpour@CERN.ch}

\affiliation {
$^{1}$School of Particles and Accelerators, Institute for Research in Fundamental Sciences (IPM), P.O.Box 19395-5531, Tehran, Iran \\
$^{2}$Department of Physics, University of Science and Technology of Mazandaran, P.O.Box 48518-78195, Behshahr, Iran
}

\date{\today}

%
%%%%%%%%%%%%%%%%%%%%%%%%%%%%%%%%%%%%%%%%%%%%%%%%%%%%%%%%%%%%%%%%%%%%%%%%%%%%%%%%%%%%%%%%%%%%%%%%%%%%%%%
\begin{abstract}\label{abstract}

The main aim of this paper is a new determination of transverse momentum dependence of unpolarized fragmentation function (TMD FFs) in single inclusive hadron production in electron-positron annihilation (SIA) process. Motivated by the need for a reliable and consistent determination of TMD FFs, we use the most recent TMD production cross sections of charged pions ($\pi^\pm$), kaons ($K^\pm$) and protons/antiprotons ($p/\bar{p}$) measured in inclusive $e^+e^-$ collisions by Belle Collaboration. These datasets are the first transverse momentum dependence of identified light charged hadron measurements SIA process. In this analysis, referred to as {\tt SK19 TMD FFs}, the common Gaussian distribution is used for the $P_{hT}$ dependent of the cross section. The uncertainties in the extraction of {\tt SK19 TMD FFs} are estimated using the standard ``Hessian'' technique. We study the quality of the TMD FFs determined in this analysis by comparing with the available recent Belle cross sections measurement. For all hadron species, we found a very good agreement between this particular set of experimental data and the corresponding theory calculations over a relatively wide range of transverse momentum $P_{hT}$. As a result of this study, suggestions are  identified for possible future research considering the theory improvements and other available experimental observables.

\end{abstract}
%

%\pacs{12.39.-x, 14.65.Bt, 12.38.-t, 12.38.Bx}

\maketitle
%\tableofcontents{}

%======================================================================
\section{Introduction}\label{sec:intro}
%======================================================================

Understanding the structure of hadron based on their fundamental particles, quarks and gluons, is gained substantial interest 
for the theoretical and experimental high energy physics communities. 
Mostly, the significant information on the hadron structure in terms of their constituents is provided from the high energy charged lepton-nucleon ($\ell N$) deep inelastic scattering (DIS) experiments at HERA collider for a wide kinematic range of momentum fraction $x$~\cite{South:2016cmx,Aguilar:2019teb}.   
In addition to the HERA experiment, the large hadron collider (LHC) is also providing valuable information in TeV scale hadron-hadron ($pp$) collisions.  

The high energy particle physics community is preparing for the extensive precision at Run III of the LHC working with the luminosity a factor of five greater than the LHC~\cite{AbdulKhalek:2019mps}. The rich physics prospects is expected at the so-called High-Luminosity LHC (HL-LHC)~\cite{Atlas:2019qfx}. The expected  precision in HL-LHC have recently been discussed in detail considering physics within and beyond the Standard Model (SM)~\cite{Azzi:2019yne,CidVidal:2018eel} and Higgs physics~\cite{Cepeda:2019klc}.

In QCD, a precise determination of the gluon and quark structure of the nucleon which entitled as the parton distribution functions (PDFs), is an essential ingredient for the theory predictions in DIS experiments at HERA and the hadron-hadron collisions at TEVATRON and LHC. In addition to the PDFs, the transverse momentum dependent parton distribution functions (TMD PDFs) and fragmentation functions (TMD FFs) are necessary ingredients for this aim and also became important topics in high-energy spin physics. 

TMD PDFs describe the densities of quarks and gluons carrying the momentum fraction $x$ of nucleon momentum by considering the spin and angular momentum properties. The TMD FFs provides deeper insight on the hadronization processes, where a hadron $h$ carrying the momentum fraction $z$ of the fragmenting parton and depends on the hadron's transverse momentum $P_{hT}$.
Semi Inclusive Deep Inelastic Scattering (SIDIS) processes are mostly used to study the TMD PDFs and TMD FFs in which they couple together in the physical observables~\cite{Radici:2018iag,Bacchetta:2017gcc,Anselmino:2015sxa,Kang:2015msa,Anselmino:2014pea,Echevarria:2016scs,Bertone:2019nxa}. One can find the most famous TMDs in SIDIS processes which called Sivers function~\cite{Sivers:1989cc,Sivers:1990fh} and the Collins FFs~\cite{Collins:1992kk}. The SIDIS process measurements to calculate the TMD effects are reported by the HERMES Collaboration~\cite{Airapetian:2004tw,Airapetian:2010ds} at HERA, the COMPASS Collaboration~\cite{Adolph:2012sn}  at CERN, and the JLab HALL A  high luminosity experiments~\cite{Qian:2011py}.

Another process which is commonly used to calculate the TMD FFs is single or double inclusive electron positron annihilations. 
As in the transverse-momentum integrated FFs, the cleanest process can be achieved in $e^+e^-$ annihilation because there is no contribution from transverse dependence of PDFs.
Belle, BABAR and BESIII Collaborations have reported the azimuthal angular asymmetries of two hadron productions in electron positron annihilation processes ($e^+e^- \rightarrow h_1 h_2 X$) at $\sqrt{s}= 10.52$~GeV, $\sqrt{s}= 10.6$~GeV and $\sqrt{s}= 3.65$~GeV, respectively~\cite{Abe:2005zx,Seidl:2008xc,TheBABAR:2013yha,Ablikim:2015pta}.  Recently, the BABAR Collaboration at SLAC has published the measurements which can be used to extract the polarized TMD FFs~\cite{TheBABAR:2013yha,Aubert:2015hha}. Due to the lack of the experimental information for the case of unpolarized TMD FFs, they can not well determined. 

Some old datasets for single unidentified light charged hadron productions in electron-positron annihilation have been presented in TASSO collaboration~\cite{Althoff:1983ew,Braunschweig:1990yd}. Most recently, these datasets have been included by a theoretical collaboration and they have extracted the TMD FFs into the unidentified light charged hadrons in single inclusive electron positron annihilation (SIA)~\cite{Boglione:2017jlh}. Thanks to the Belle collaboration which has provided the first measurements of the production unpolarized cross sections of pions, kaons, as well as protons in SIA process at $10.58$ GeV at B-factories~\cite{Seidl:2019jei}. These observables are as a function of three variables: the parton fractional energy carried by hadron $z$, the event-shape variable called thrust $T$, and the hadron transverse momentum with respect to the thrust axis~\cite{Seidl:2019jei}. These measurements can be used for studying the transverse momentum dependence of unpolarized single hadron FFs $D(z, P_{hT}, Q)$ and also obtaining a better theoretical predictions for the various TMDs in transverse spin asymmetries in SIDIS, electron-positron annihilation and proton-proton collisions.

Motivated by this need for a reliable and consistent determination of transverse momentum dependence of unpolarized fragmentation functions (TMD FFs) into pion, kaon and proton and their uncertainties, we present in this work a first TMD FFs analysis, entitled  {\tt SK19 TMD FFs}, based on the most recent Belle measurements in single production of these three identified light charged hadrons in electron positron annihilation processes~\cite{Seidl:2019jei}. 
We will present an interesting results from Gaussian parametrization of the TMD FFs in which the parametrizations depend on both momentum fraction $z$ and transverse momentum $P_{hT}$. We show that an accurate predictions can be obtained considering the features of TMD factorization and evolution in the non-perturbative QCD. 
The {\tt SK19 TMD FFs} sets are constructed following the general fitting methodology outlined in previous {\tt SGK FFs} studies~\cite{Soleymaninia:2018uiv,Soleymaninia:2019sjo}, which utilizes ``Hessian'' techniques to obtain a faithful estimate of TMD FFs uncertainties. Together with several other improvements, we present a validation of the {\tt SK19 TMD FFs} results through detailed comparison with the analyzed datasets. We show that the theory predictions based on {\tt SK19 TMD FFs} are in agreement with the Belle measurements over the wide range of $z$ and $P_{hT}$.  
	
The outline of this paper is as follows: In Sec.~\ref{sec:exper} we introduce in details the recent Belle Collaboration datasets for pion, kaon and proton in SIA and different kinematical cuts for the various hadrons. Then, in Sec.~\ref{sec:theory}, we discuss the theoretical framework for  definition of TMDs in terms of factorization theorems and TMD evolution equation along with the TMD of single-hadron production in electron-positron annihilation. Sec.~\ref{sec:Fitting} contains a detailed discussions of our parametrization for the TMD part of the pion, kaon and proton FFs. The description of the fitting strategy, including the minimization procedure, the choice of parametrization and the estimation of uncertainties associated with the TMD FFs is presented in this section. The obtained results are clearly discussed in Sec.~\ref{sec:results} for various hadrons and our theory predictions based on the extracted pion, kaon and proton TMD FFs are compared with the Belle cross section data. Lastly, we conclude with a summary in Sec.\ref{sec:conc}.

%======================================================================
\section{Experimental data}\label{sec:exper}
%======================================================================

Very recently Belle Collaboration at KEKB has published the measurements of the single production cross sections $(d^3\sigma /dzdP_{hT}dT)$ of charged pion, kaon and proton/antiproton as a function of hadron fractional energy $z$, the event-shape variable (Thrust) and the transverse momentum $P_{hT}$ at the center-of-mass energy of $\sqrt{s}=10.58$~GeV~\cite{Seidl:2019jei}. The thrust variable, $T$, is related to the thrust axis $\hat{n}$. Experimentally, the datasets depended on the transverse momentum are calculated relative to the thrust axis $\hat{n}$ and the event-shape variable thrust $T$ is the maximum of the following equation
%------------------------------------------------------
\begin{equation}\label{thrust}
T^{\rm max} = \frac{\sum _h {\mathbf P}_h^{\rm CMS}.
\hat{{\mathbf n}}}{\sum _h {\mathbf P}_h^{\rm CMS}}.
\end{equation}
%------------------------------------------------------
The ${\mathbf P}_h^{\rm CMS}$ denotes the momentum of hadron $h$ in the center-of-mass energy framework (CMS).  Hence, the datasets in this experiment are presented in the bins of the thrust value, $T$. 
The Belle Collaboration at KEKB has reported the hadron cross sections for charged pion, kaon and proton/antiproton at 18 equidistant $z$ bins in [0.1~-~1.0] region, 20 equidistant $P_{hT}$ bins in [0.0~-~2.5] region and 6 thrust bins with $0.5, 0.7, 0.8, 0.85, 0.9, 0.95$ and $1$ boundaries. 
Belle Collaboration has presented the distributions of thrust for different hadrons species in their measurement~\cite{Seidl:2019jei}. It has been shown in Ref.~\cite{Seidl:2019jei} that the $uds$- and charm events peak at thrust range of $0.85 < T< 0.9$ showing that at high thrust values their contributions becomes sizable. For this reason,  Belle Collaboration has analyzed and presented the results for this specific range of $T$. In our analysis, we also follow the analysis by Belle Collaboration and focus on the datasets for this particular thrust bin and present our results for the trust range of $0.85 < T < 0.9$.

Due to the large uncertainties associated with the differential cross sections at large $z$ regions, we exclude some of high-$z$  data  points from our TMD FFs QCD analysis. 
Moreover, since the low values of $z$ can not be applicable for factorization theorem and in order to get the reliable fits, the kinematical cuts on small $z$ region are also imposed to the datasets. The kinematical cuts on $z$ are different for various hadrons analyzed in this study. The details of the kinematical cut applied on the datasets are reported in Table.~\ref{tab:datasets} as well.

According to the factorization theorem the differential cross section which depends on the fractional energy $z$ and transverse momentum $P_{hT}$ is written as follows, 
%------------------------------------------------------
\begin{eqnarray}\label{difcross+tansora}
\frac{d\sigma ^h}{dz\,dP_{hT}} =L_{\mu \nu}(W^{\mu \nu}_{\rm TMD}+W^{\mu \nu}_{\rm coll}),
\end{eqnarray}
%------------------------------------------------------
while $L_{\mu \nu}$ is the leptonic tensor and $W^{\mu \nu}_{\rm TMD}$ and $W^{\mu \nu}_{\rm coll}$ are the hadronic tensors. The first hadronic tensor $W^{\mu \nu}_{\rm TMD}$ has contribution in the region of small transverse momenta, while the second one $W^{\mu \nu}_{\rm coll}$ contains collinear factorization.  

Generally speaking, in some certain regions with $P_{hT} \sim 2$ GeV, the collinear contributions in the cross section are more than TMD contributions, while for the kinematical region of $P_{hT} < 1$ GeV the TMD term has the largest contribution in the cross section. In addition, one could try to perform an analysis in the non-perturbative evolution region, and hence, for this purpose the $P_{hT}$ should be restricted to the $P_{hT} < 1$ GeV. We follow this assumption to perform our QCD analysis. We should mention here that the uncertainties of observables for the individual $z$ bins increase for the high-$P_{hT}$ value.
Hence, we exclude the experimental data with $P_{hT} > 1$ GeV for different identified light charged hadron fits while the range of data analyzed for proton/antiproton are wider than charged pion and kaon. 

In order to finalize the maximum cut on $P_{hT}$, the sensitivity of $\chi^2$ to the variations of $P_{hT} < {P^{\rm max}_{hT}}$ is investigated for the TMD dependence of SIA data from Belle Collaboration. We scan the $P_{hT}$ region of  $0.3 < {P^{\rm max}_{hT}} <1.1$ GeV for the pion, kaon and proton TMD FFs analyses. Considering these $\chi^2$ scans, our TMD FFs fits are presented for each different $P_{hT} < {P^{\rm max}_{hT}}$ cut.

%
%--------------------------------
%
\begin{figure*}[htb]
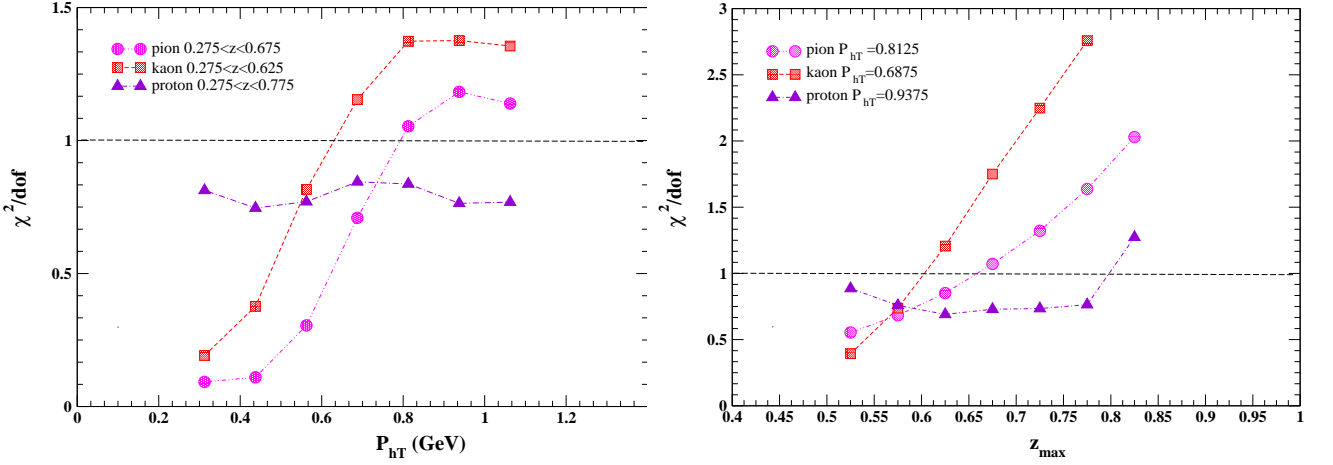

	\begin{center}
		\vspace{0.5cm}
		%\hspace{-2mm}}
		\resizebox{0.480\textwidth}{!}{\includegraphics{chidof_0-08.eps}}
		\resizebox{0.480\textwidth}{!}{\includegraphics{chidof_0-08_zmax.eps}}
		\caption{ Dependence of $\chi^2/{\rm dof}$ on the maximum cut values of ${P_{hT}}$ and $z$ for  $0.3 < {P^{\rm max}_{hT}} <1.1$ GeV  and $0.5 < {z_{\rm max}} <0.85$ datasets of pion, kaon and proton used in the analyses. } \label{fig:chidof}
	\end{center}
\end{figure*}
%
%--------------------------------
%

In Fig.~\ref{fig:chidof}, the dependence of $\chi^2/{\rm dof}$ to the  maximum cut value of $P_{hT}$ has been presented for pion, kaon and proton/antiproton. 
As one can conclude from the figure, the best $\chi^2/{\rm dof}$ value for the pion is related to the fit to the data with ${P^{\rm max}_{hT}} = 0.9$~GeV. As can be seen from Fig.~\ref{fig:chidof} there is no further improvement on the $\chi^2/{\rm dof}$ for the values larger than ${P^{\rm max}_{hT}} > 0.9$ GeV. 
Our investigations for the ${P^{\rm max}_{hT}}$ dependence for the kaon and proton/antiproton reflect different findings. As Fig.~\ref{fig:chidof} clearly shows, the best high-$P_{hT}$ cut for the kaon and proton/antiproton need to be taken as ${P^{\rm max}_{hT}}=0.8$ and ${P^{\rm max}_{hT}}=1$ GeV, respectively.

Since TMD FFs also depend on $z$ parameter, it would be interesting to investigate the sensitivity of $\chi^2$ to the particular value of  $z_{\rm max}$ and exclude the datasets with $z > z_{\rm max}$ from the analysis. The $\chi^2/{\rm dof}$ as a function of the maximum value of $z$ is shown in Fig.~\ref{fig:chidof} for different hadrons. Note that in these scans the optimal values for ${P^{\rm max}_{hT}}$ obtained for the different hadron species are considered.
As can be seen  from  Fig.~\ref{fig:chidof}, one can conclud that the best values of $\chi^2/{\rm dof}$ in respect to the particular value of $z$ are different for pion, kaon and proton/antiproton. For the pion we obtained $z_{\rm max}=0.675$ and there is no further improvement on adding the datasets with larger values than $z_{\rm max}=0.625$ and $z_{\rm max}=0.775$  for kaon and proton/antiproton, respectively. 

%
%-------------------------------------------------------------------------------
%
\begin{table*}[htb]
\renewcommand{\arraystretch}{2}
\centering 	\scriptsize
\begin{tabular}{lccccr}				\hline
Hadron        ~&~  $z$ cut ~&~  $P_{hT}$ cut    ~&~  data points~ & ~$\chi^2/{\rm dof}$ \\
\hline \hline
$\pi^\pm$   &  [0.275~-~0.675]    & [0~-~0.9]     & 63 & 1.053  \\
$K^\pm$     &  [0.275~-~0.625]    & [0~-~ 0.8]  & 48 & 1.154   \\
$p/\bar{p}$ &  [0.275~-~0.775]    & [0~-~1]   & 88 & 0.755   \\ 				
\hline \hline	
\end{tabular}
\caption{ \small The input datasets included in the three individual analyses for $\pi^\pm$, $K^\pm$ and $p/\bar{p}$.  For each hadron, we indicate the kinematical cuts of $z$ and $P_{hT}$, number of data points in the fits and the $\chi^2/{\rm dof}$ values for each datasets. }
\label{tab:datasets}
\end{table*}
%
%-------------------------------------------------------------------------------
%

The summary of our final data selection considering the $z$ and $P_{hT}$ kinematical cuts applied for various hadrons are presented in Table.~\ref{tab:datasets}.  The second and third columns of this table show our choice of $z$ and $P_{hT}$, respectively.  In the fourth column of this table, the number of data points for every hadron is presented. Finally, we reported the $\chi^2/{\rm dof}$ determined from the fits of pion, kaon and proton/antiproton, respectively. The details of our TMD FFs parametrization and fitting methodology to calculate the TMD FFs for light hadrons and $\chi^2$ will be discussed in Sec.~\ref{sec:Fitting}.

%======================================================================
\section{ Factorization framework and TMD FFs }\label{sec:theory}
%======================================================================

 The factorization theorems for perturbative QCD (pQCD) are key instruments in QCD phenomenology and hadron structure. 
There are two kinds of factorizations that have been extensively used in QCD analyses which are Collinear and TMD factorizations. In Collinear factorization the FFs depend only on the longitudinal momentum fraction and the transverse momentum components are integrated over. In TMD factorization, the FFs depend on transverse momentum in addition to the momentum fraction variable.
A TMD factorization formalism is constructed by Collins, Soper and Sterman (CSS) and they provided a systematic formalism of pQCD all over the range of transverse momentum. The details of such factorization are fully discussed in Refs.~\cite{CSS1,CSS2,CSS3}. The detailed explanations of such calculations is out of scope of the following paper. Hence, here we briefly review the important points and the main features of this formalism in this section. The TMD FFs energy evolution is given by
%------------------------------------------------------
\begin{eqnarray}
\label{eq:difcross}
\frac{d \ln 
	\tilde{D} (z,{\bf b}_T,
	 \mu,\zeta_D)}
 {d \ln \mu} =
  \gamma_D(g(\mu),
   \zeta_D /\mu^2)\,.
\end{eqnarray}
%------------------------------------------------------
By considering $k_T$ as a transverse component of the momentum of fragmenting parton to the final hadron, $b_T $ refers to the conjugate variable to $k_T$.
%------------------------------------------------------
\begin{multline}
\label{eqFFTMD}
\tilde{D}_{h/f}(z,{\bf b}_T,
 \mu,\zeta_D) = \\
\sum_j \int_z^1 
\frac{d \hat{z}}{\hat{z}^{3-2\epsilon}}
 \tilde{C}_{j/f}(z/\hat{z},
 b_{\ast}, \mu_b^2,
 \mu_b,g(\mu_b))
  D_{h/j}(\hat{z},
  \mu_b) \\
\times e^ { \ln 
	\frac{\sqrt{\zeta_D}}{\mu_b}
 \tilde{K}(b_{\ast},
  \mu_b) + 
\int_{\mu_b}^\mu 
\frac{d \mu^\prime}{\mu^\prime}
 [ \gamma_D(g(\mu^\prime),1) 
- \ln \frac{\sqrt{\zeta_D}}
{\mu^\prime}
 \gamma_K(g(\mu^\prime)) ] } \\
\times 
e^ { g_{H/j}(z,b_T) + g_K(b_T)
 \ln \frac{\sqrt{\zeta_D}}
 {\sqrt{\zeta_{D,0}}} }.
\end{multline}
%------------------------------------------------------
In above equation, the $\mu$ is the renormalization scale, $\zeta_D$ is a regulator for light-cone divergences, $\tilde{C}_{j/f}$ are the Wilson coefficients, 
$\gamma_D$ and $\gamma_K$ are anomalous dimensions and $\tilde{K}$ is the Collins-Soper (CS) kernel.
The details component of Eq.~\ref{eqFFTMD} can be found in literature and we refer the reader to the Ref.~\cite{Collins} for clear review.

The $\tilde{C}_{j/f}(z/\hat{z},b_{\ast}, \mu_b^2,\mu_b,g(\mu_b))$, $\tilde{K}(b_{\ast}, \mu_b)$, $\gamma_D(g(\mu^\prime),1)$ and $\gamma_K(g(\mu^\prime)$ are perturbative functions which calculable in perturbative QCD. At the first line of Eq.~\eqref{eqFFTMD}, $D_{h/j}(\hat{z}, \mu_b)$ indicates to the integrated FFs from collinear factorization which is a non-perturbative function. The functions $g_{H/j}(z,b_T)$ and $g_K(b_T)$ are also non-perturbative quantities. At lowest order of the coefficient functions, the first sentence is simply $D(z,\mu_b)$.  By considering the usual choices for arbitrary quantities $\mu \rightarrow Q$, $\zeta_D\rightarrow Q^2$ and
$\zeta_D^{(0)} \rightarrow Q_0^2$, the CS kernel $\tilde{K}(b_{\ast}; \mu_b)$ vanishes at order of $\alpha_s$.
 
Finally, we are interested in the momentum space TMD definition ($zk_T$ as the transverse momentum parameter). It defines as the Fourier transform of coordinate space for TMD FFs,
%------------------------------------------------------
\begin{eqnarray}\label{FFTMD}
&&D_{h/f}(z,z  {\bf k}_T,
\mu,\zeta_D) = \nonumber \\
&&\frac{1}{(2 \pi)^2} 
\int d^2 {\bf b}_T \,
e^{-i {\bf k}_T \cdot
	 {\bf b}_T} \, 
\tilde{D}_{h/f}(z,
 {b},\mu,\zeta_D).
\end{eqnarray}
%------------------------------------------------------
We assume the dependence on the $z$ and the magnitude of the transverse momentum $k_T$ to be factorized and at the initial scale, we assume a Gaussian form for the $k_T$ dependence of TMD FFs which is given by,

%------------------------------------------------------
\begin{eqnarray}\label{Gaussian}
D_{h/f}(z,k_T) = 
D_{h/f}(z)
\times \frac{e^{ -k_T^2 /
		 \langle 
k_T^2\rangle}}{\pi 
\langle k_T^2 \rangle},
\end{eqnarray}
%------------------------------------------------------
 
According to the factorization of TMD, the differential cross section for $e^+e^- \rightarrow h X$  process, which there is one observed hadron at the final state, can be determined considering the following equation~\cite{Boglione:2017jlh},

%------------------------------------------------------
\begin{eqnarray}\label{eq:difcross}
\frac{d \sigma ^h}
{dz\,dP_{hT}} &=& 
2 \pi P_{hT} \frac{4
	\pi \alpha ^2}{3s} 
\sum \limits_q 
{{e_q}^2}\, {\cal N }
D_{h/f}(z,{P_{hT}},
 {Q^2})  \nonumber \\ 
&=&2\pi P_{hT}
\sigma_{\rm tot}
\sum \limits_q  {\cal N }
D_{h/f}(z,Q^2) \;
 h^h(P_{hT})\,,
\end{eqnarray}
%------------------------------------------------------

which is expressed at the leading order ($\alpha_s$), and $Q$  in Eq.~\eqref{eq:difcross} is the hard scale of the process. The total inclusive cross section $\sigma_{\rm tot}$ at leading order is given by,

%------------------------------------------------------
\begin{eqnarray}\label{eq:totalcross}
\sigma_{\rm tot} = \sum \limits_q {{e_q}^2}\, \frac{4 \pi \alpha ^2}{3s} \,.
\end{eqnarray}
%------------------------------------------------------

It should be note here that, in order to give more flexibility to the TMD distributions, we defined an appropriate normalization parameter ${\cal N }$.   This normalization parameter will be calculated from the fit. Then we keep it fixed on it's best fit value in the final minimization to calculate the free parameters of TMD fragmentation function.  
The sum is over all quark and anti-quark flavors and hence the gluon does not have contribution at this accuracy. We refer the reader to Ref.~\cite{Collins,Aybat:2011zv} for the detailed discussions.

The TMD fragmentation function can be expressed considering two terms. The first term is the  unpolarized collinear FF $D_{h/f}(z, Q^2)$ and  the second term corresponds to the TMD dependent $h^h(P_{hT})$ which is not dependent on the scale of energy and also the flavor.  Consequently, the TMD FFs can be given by, 
%------------------------------------------------------
\begin{eqnarray}\label{eq:D_q}
D_{h/f}(z,{P_{hT}}, Q^2)=  D_{h/f}(z,Q^2) \; h^h(P_{hT})\,.
\end{eqnarray}
%------------------------------------------------------
Over the past decades,  many studies are performed for the determination of unpolarized FFs by including experimental data from different processes such as SIA, SIDIS and hadron-hadron collisions. The most recent calculations for the unpolarized FFs for various hadrons and at different  QCD accuracies can be  found in Refs.~\cite{Soleymaninia:2019sjo,Epele:2018ewr,Bertone:2017tyb,Soleymaninia:2018uiv,Soleymaninia:2017xhc,deFlorian:2014xna,deFlorian:2017lwf,Bertone:2018ecm,Sato:2016wqj,Nocera:2017gbk}. 
For the unpolarized FFs in Eq.~\eqref{eq:D_q}, we use the most recent analysis of pion, kaon and proton FFs by {\tt NNFF1.0} Collaboration~\cite{Bertone:2017tyb}. It should be noted here that the $h(P_{hT})$ in Eq.~\eqref{eq:D_q} only depends on the $P_{hT}$. As we mentioned earlier, our theory calculations are limited to the LO perturbative QCD.  The {\tt NNFF1.0} Collaboration determined the unpolarized FFs for charged pion, kaon and proton/antiproton data by including SIA experimental data up to the NNLO accuracy. For the purpose of our QCD analysis, we use the LO FFs from {\tt NNFF1.0} Collaboration.

In the next section, the {\tt SK19 TMD FFs } parametrization for the $D_{h/f}(z,{P_{hT}}, Q^2)$, the fitting methodology and the minimization process will be discussed in details.

%======================================================================
\section{ Fitting methodology and {\tt SK19 TMD FFs } parametrization }\label{sec:Fitting}
%======================================================================

In this section, we describe the fitting methodology and {\tt SK19 TMD FFs } parametrization applied in this analysis for the determination of TMD of charged pion, kaon and proton/antiproton. The methodology presented here follows from the standard QCD analyses, however a number of improvements have been implemented in this work.

First, we discuss the details of the analysis, together with the framework that need to be considered in order to deal with the determination of the TMD FFs, such as the parameterization of the TMD FFs at the input scale. Then, following the SGK-FFs methodology~\cite{Soleymaninia:2019sjo,Soleymaninia:2018uiv}, we present the method of the minimization and the uncertainties associated with the TMD FFs which are estimated using the standard Hessian approach.

Now we are in a position to present the {\tt SK19 TMD FFs } parametrization form for the phenomenological study of the transverse momentum distributions. Following the analyses of Refs.~\cite{Boglione:2017jlh,Anselmino:2013lza,Anselmino:2018psi} we use the Gaussian form which is the most commonly parametrization for TMD FFs and widely used in the QCD analysis of Drell-Yan, SIDIS and also SIA processes,
%------------------------------------------------------
\begin{equation}
D_{h/f}(z,{P_{hT}}, Q^2) =
D_{h/f}(z,Q^2) \frac{{e^ {- {P_{hT} }^2/
			\langle {{P_{hT} }^2} \rangle }}} 
{{\pi \langle {{P_{hT} }^2} \rangle }} \,.
\label{eq:gauss}
\end{equation}
%------------------------------------------------------
Since the SIA datasets published by Belle experimental depend on the $z$ parameter other than the transverse momentum, a $z$- dependent of $\langle {{P_{hT} }^2 \rangle}$ may be appropriate. Then we define the following functional form,
%------------------------------------------------------
\begin{equation}
\label{eq:widthzdep}
\langle P_{hT}^2 \rangle = \alpha + z^\beta (1 - z)^\gamma ,
\end{equation}
%------------------------------------------------------
which $\alpha$, $\beta$ and $\gamma$ are the free parameters and need to be determined from QCD fit to the Belle experimental data. The small and large region of momentum fraction $z$, will be controlled by the parameters $\beta$ and $\gamma $, respectively. Accordingly, there are 4 unknown parameters including normalization factor ${\cal N }$ which provide enough flexibility to have a reliable fit.
% Since $\langle {{P_{hT} }^2 \rangle}$ depends on $z$, the TMD dependent part of FF $h$ will depend on $P_{hT}$ and also $z$ in our parametrization model.

We start now by briefly reviewing the standard minimization procedure. To determine the best fit in our TMD FFs analysis, one needs to minimize the $\chi^2$ function with the free unknown parameters presented in Eq.~\eqref{eq:widthzdep} together with the normalization factor ${\cal N }$ in Eq.~\eqref{eq:difcross}. Likewise all QCD analysis, the $\chi^2(\rm p)$ quantifies the goodness of the fit to the datasets for a set of independent parameters ${\rm p}$ that specifies the TMD FFs at the input scale, Q$_0$ = 5 GeV. This standard $\chi^2 ({\rm p})$ function is expressed as,
%---------------------------------------------
%
\begin{equation}
\label{eq:chi2}
\chi_n^2 (\rm p) = 
\sum_{i=1}^{N_n^{\rm data}}
\left(\frac{ {\cal E}_i -
{\cal T}_i (p) }{\, 
\Delta ({\cal E}_i)}
\right)^2\,.
\end{equation}
%
%---------------------------------------------
In above $\chi_n^2 (\rm p)$ function, ${\cal E}$, ${\cal T}$  and $\Delta ({\cal E}_i)$ indicate the experimental measurement, the theoretical value for the $i^{\rm th}$ data point, and the experimental uncertainty (statistical and systematic combined in quadrature), respectively. In our TMD FFs analysis, the minimization of the above $\chi^2 (\rm p)$ function has been done using the CERN program library {\tt MINUIT}~\cite{James:1994vla}. The normalization factor $\cal N$ appears as free parameters in the fit. It is determined simultaneously with the fit parameters of the functional forms of Eq.~\eqref{eq:widthzdep} and then keep fixed at it's best fit value. In our analysis, we find the normalization factor $\cal N$ provides additional flexibility to achieve a good description of data. 

%---------------------------------------------
%
\begin{table}
\begin{tabular}{lcccc}
\hline
Parameters&$\pi^\pm$ ~&~ $K^\pm$ ~ & ~ $p/\bar{p}$    \\ 
		\hline \hline
		$\alpha$ &      $0.105$     & $0.002$       &  $ 0.240$        \\
		$\beta$ &      $1.413$     &  $1.077$       &   $4.648$        	\\
		$\gamma$ &      $0.854$     &  $0.739$       &   $1.153$        \\
		${\cal N }$ &             $0.290^*$    &  $ 0.166^*$      &    $0.335^*$      \\ \hline
\end{tabular}
\caption{ The best-fit parameters for the {\tt SK19 TMD FFs } into  $\pi^\pm$,  $k^\pm$  and $p/\bar{p}$. The values labeled by (*) have been fixed. The details
	of the determination of best fit values are described in the text. }
\label{table:pars}
\end{table}
%
%---------------------------------------------

The results of our fit are shown in Table.~\ref{table:pars} for charged pion, kaon and proton/antiproton separately.
We start to determine all shape parameters of Eq.~\eqref{eq:widthzdep} for the TMD FFs of $\pi^\pm$, $K^\pm$ and $p/\bar{p}$ from fit to Belle datasets. 

For calculating the uncertainties of the TMD FFs, we follow our previous analyses and use the standard ``Hessian" method~\cite{Pumplin:2001ct,Martin:2009iq,deFlorian:2011fp,Schmidt:2018hvu,Eskola:2009uj} with $\Delta \chi^2 = 1$ at 68\% confidence level (CL).
In Hessian method, the uncertainty on $\langle P_{hT}^2 \rangle(z)$ in Eq.~\eqref{eq:widthzdep} can be obtained from linear error propagation,
%--------------------------------
\begin{eqnarray}
\label{uncertainties}
[\Delta \langle P_{hT}^2 \rangle]^2 = && \nonumber \\
\Delta \chi^2_{\mathrm global} \times \,   
&& \big [ \sum_i (\frac{\partial \Delta \langle P_{hT}^2 \rangle(z,
{\hat \eta})}{\partial \eta_i})^2 \,  C_{i i}
 +  \nonumber \\
&&\sum_{i \neq j}
( \frac{\partial \Delta  \langle P_{hT}^2 \rangle(z, {\hat \eta})}
{\partial \eta_i} 
\frac{\partial \Delta  \langle P_{hT}^2 \rangle(z,
{\hat \eta})}{\partial \eta_j} )
\,  C_{i j}  \big ]\,,  \nonumber \\
\end{eqnarray}
%--------------------------------
where $\eta_i$ ($i$ = 1, 2, ..., N) denotes to the free parameters for TMD FFs presented in Eq.~\eqref{eq:widthzdep}.
N (=4) is the number of optimized parameters and ${\hat \eta}_i$ is the optimized parameter. $C \equiv H_{i, j}^{-1}$ in Eq.~\eqref{uncertainties} indicate to the elements of the covariance matrix determined using the CERN program library {\tt MINUIT}~\cite{James:1994vla} in the TMD FFs analysis at the input scale $Q_0$. The $T = \Delta \chi^2_{\rm global}$ is the tolerance for the required CL which is considered to be 1 for the  68\%. Further details on ``Hessian" method can be found in Refs.~\cite{Soleymaninia:2018uiv,Soleymaninia:2019sjo}.   

With the agreement between the Belle datasets and our theory established, we are in a position now to present the main results and findings of the {\tt SK19 TMD FFs} QCD analysis in the next section.  

%--------------------------------
\begin{figure*}[htb]
	\begin{center}
		\vspace{0.5cm}
		%\hspace{-2mm}}
		\resizebox{0.55\textwidth}{!}{\includegraphics{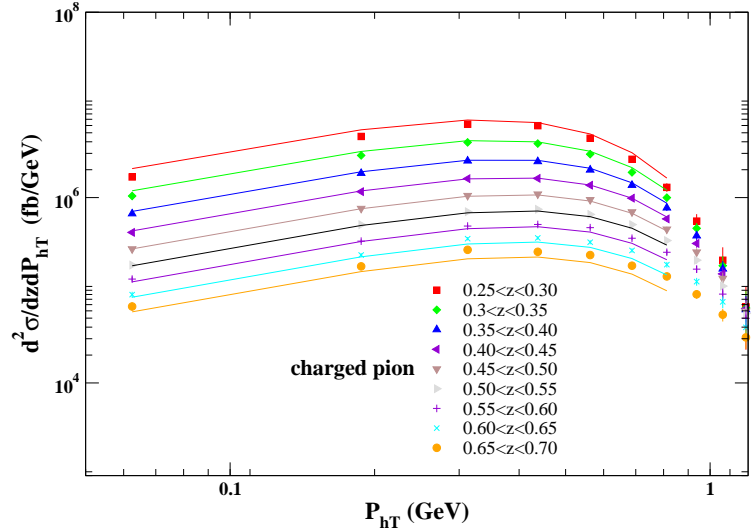}}  
		\caption{ Comparisons between the differential cross sections measurements from Belle Collaboration~\cite{Seidl:2019jei}  for pion and our theory predictions as a function of $P_{hT}$ for different $z$ bins. } \label{fig:crosspi}
	\end{center}
\end{figure*}
%--------------------------------

%--------------------------------
\begin{figure*}[htb]
	\begin{center}
		\vspace{0.50cm}
		%\hspace{-2mm}}
		\resizebox{0.55\textwidth}{!}{\includegraphics{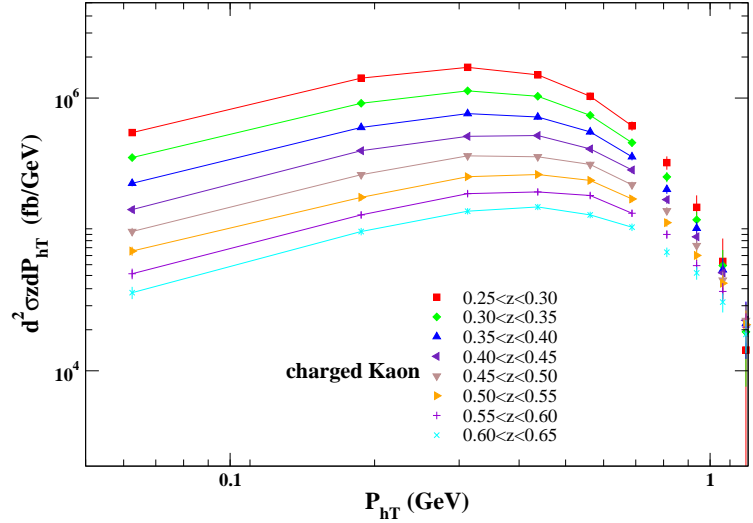}}  
		\caption{ Same as Fig.~\ref{fig:crosspi} but for kaon. } \label{fig:crosska}
	\end{center}
\end{figure*}
%--------------------------------

%--------------------------------
\begin{figure*}[htb]
	\begin{center}
		\vspace{0.50cm}
		%\hspace{-2mm}}
		\resizebox{0.55\textwidth}{!}{\includegraphics{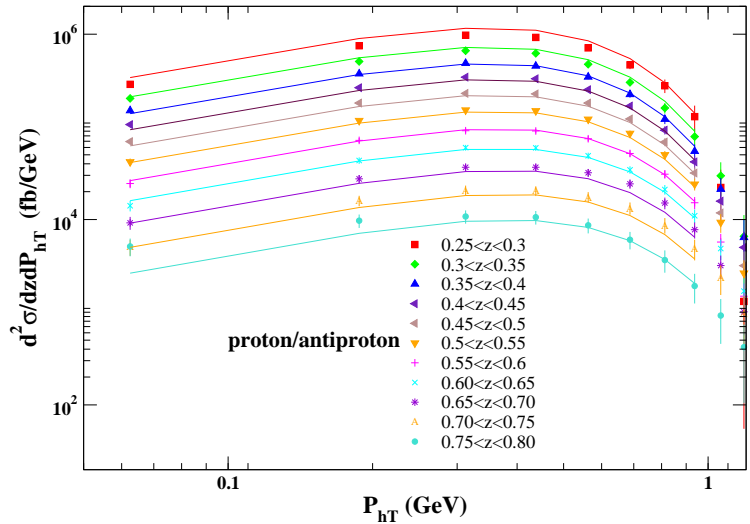}}  
		\caption{ Same as Fig.~\ref{fig:crosspi} but for proton. } \label{fig:crosslpr}
	\end{center}
\end{figure*}
%--------------------------------

%======================================================================
\section{Results  and discussions }\label{sec:results}
%======================================================================

In this section, we present the main results and findings of our analysis, namely the {\tt SK19} sets of TMD FFs. We first assess the quality of our QCD fit  which is followed by detailed discussions of the main features of {\tt SK19 TMD FFs} and comparing the resulting theory predictions with the Belle experimental data. Before moving forward, it is practical to illustrate qualitatively the expected finding for our TMD FFs analysis.

To this end, we use the unpolarized TMD FFs for pion and kaon as well as proton determined in this analysis and calculate the theoretical predictions. In Figs.~\ref{fig:crosspi}, \ref{fig:crosska} and \ref{fig:crosslpr}, the differential cross section datasets from Belle Collaboration as a function of transverse momentum $P_{hT}$ for pion, kaon and proton from SIA process are shown. These datasets are the most recent measurement from Belle  Collaboration~\cite{Seidl:2019jei}. The measured cross section presented in these figures depend also on the $z$ parameter which are indicated in different bins in the figures. The scale of energy in this determination by Belle experiment is fix for all hadron species which is equal to $10.58$ GeV.
At the level of individual datasets for each hadron spices, we find in most cases a good agreement between the theory calculations and the corresponding experimental measurements. However, the results for every hadron spices need some more discussion.

We start our discussion with the pion TMD FFs. As we mentioned in Table.~\ref{tab:datasets}, in our analysis and for the case of pion TMD FFs, the datasets are restricted to the range of  $0.275 \le z \le 0.675$ for the data points with $P_{hT} < 0.9$ GeV.  Hence, the theory prediction for the pion as shown in Fig.~\ref{fig:crosspi} are restricted to these particular values of $z$ and $P_{hT}$ kinematical cuts. As expected by the $\chi^2/{\rm dof}$ values listed in Table.~\ref{tab:datasets} for the pion analysis, the experimental measurements agree well with the theory predictions computed using the {\tt SK19 TMD FF}  sets. The agreement between the data and our theory predictions are excellent for the different bins of $z$ and the $P_{hT}$ values analyzed in this study, {\it i.e.}  $P_{hT} < 0.9$ GeV. One can see a very small shift for our theory predictions at the higher values of $z$, namely $0.65 < z < 0.70$ GeV. However, this treatment do not significantly affect our conclusion on the fit quality of pion TMD FFs analysis.

Fig.~\ref{fig:crosska} presents the comparisons between kaon experimental data and full lines of the Gaussian fits. Like for the case of pion fit, a similar argument can be made for the kaon TMD FFs fit. From $\chi^2/{\rm dof}$ values listed in Table.~\ref{tab:datasets}, one expect an excellent fit to the data. Our kaon fit are restricted to the $z$ range of $0.275 \le z \le 0.625$ and for $P_{hT} <  0.8$ GeV.  As can bee seen from Fig.~\ref{fig:crosska}, the data/theory agreements are excellent for all range of $z$ and $P_{hT}$ analyzed for the kaon TMD FFs fit. 

Our theory predictions for the  proton/antiproton as a function of $P_{hT}$ and for different bins of $z$ are shown in Fig.~\ref{fig:crosslpr}.   As one can see, the lack of agreement between the theory predictions and the data can be traced only to the high-$z$ bins, $0.75 < z < 0.80$ as well as at the small-$z$ bins, $0.25 < z < 0.30$, both for the low-$P_{hT}$, $P_{hT} < 0.30$ GeV regions of the theory predictions. Hence, one can concludes that the agreements between the theory predictions and proton/antiproton data points are less than the case of kaon TMD FFs fits. As a last point, we should mentioned here that the kinematical cuts for the proton/antiproton are wider than for the pion and kaon. As we mentioned, we analyze the data points for the proton/antiproton fit for the $z$ bins of $0.275 \le z \le 0.775$ and $P_{hT} < 1$ GeV which lead to the inclusion of more data points in this fit. 

\begin{figure*}[htb]
\vspace{0.50cm}
\resizebox{0.480\textwidth}{!}{\includegraphics{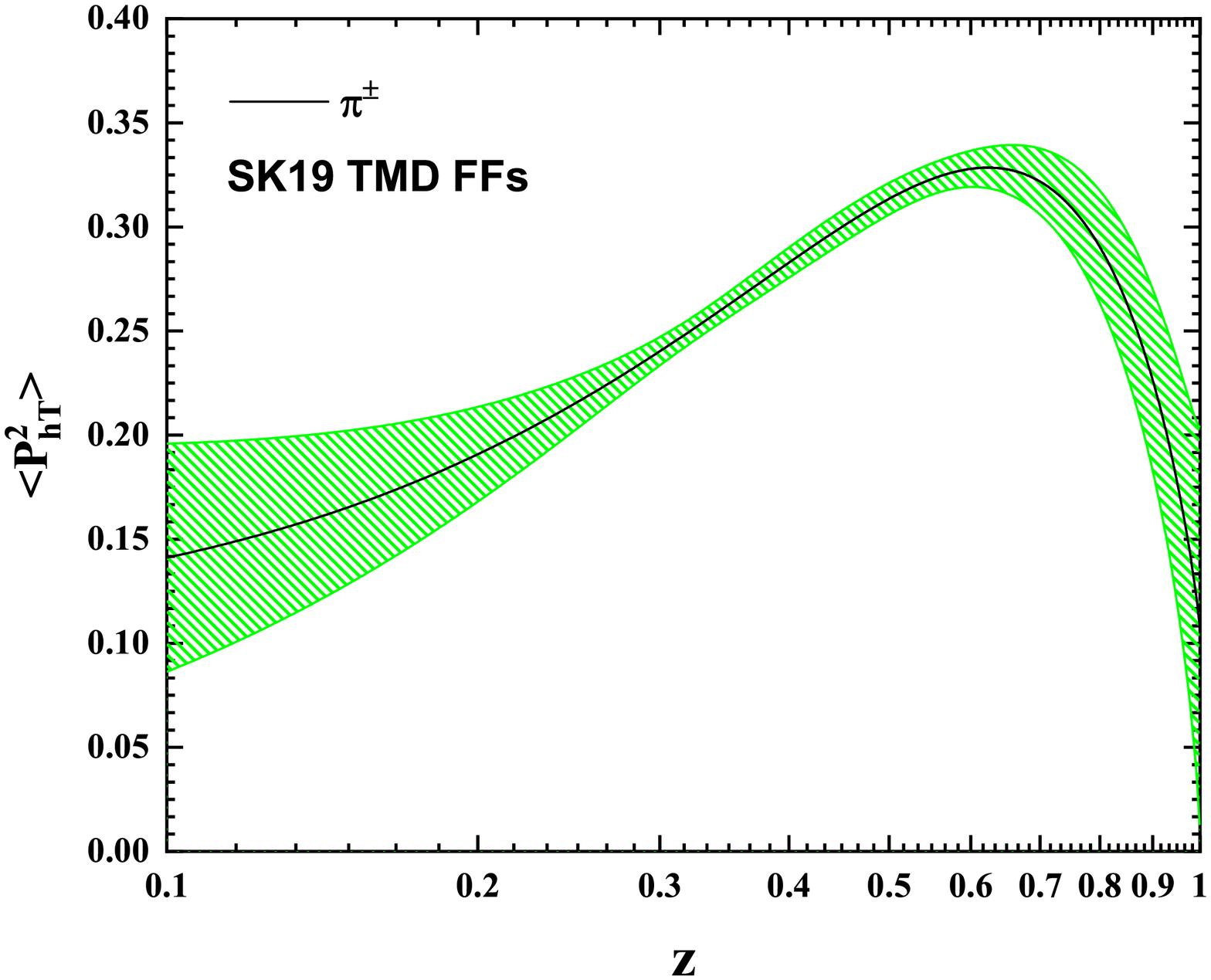}} 	
\resizebox{0.480\textwidth}{!}{\includegraphics{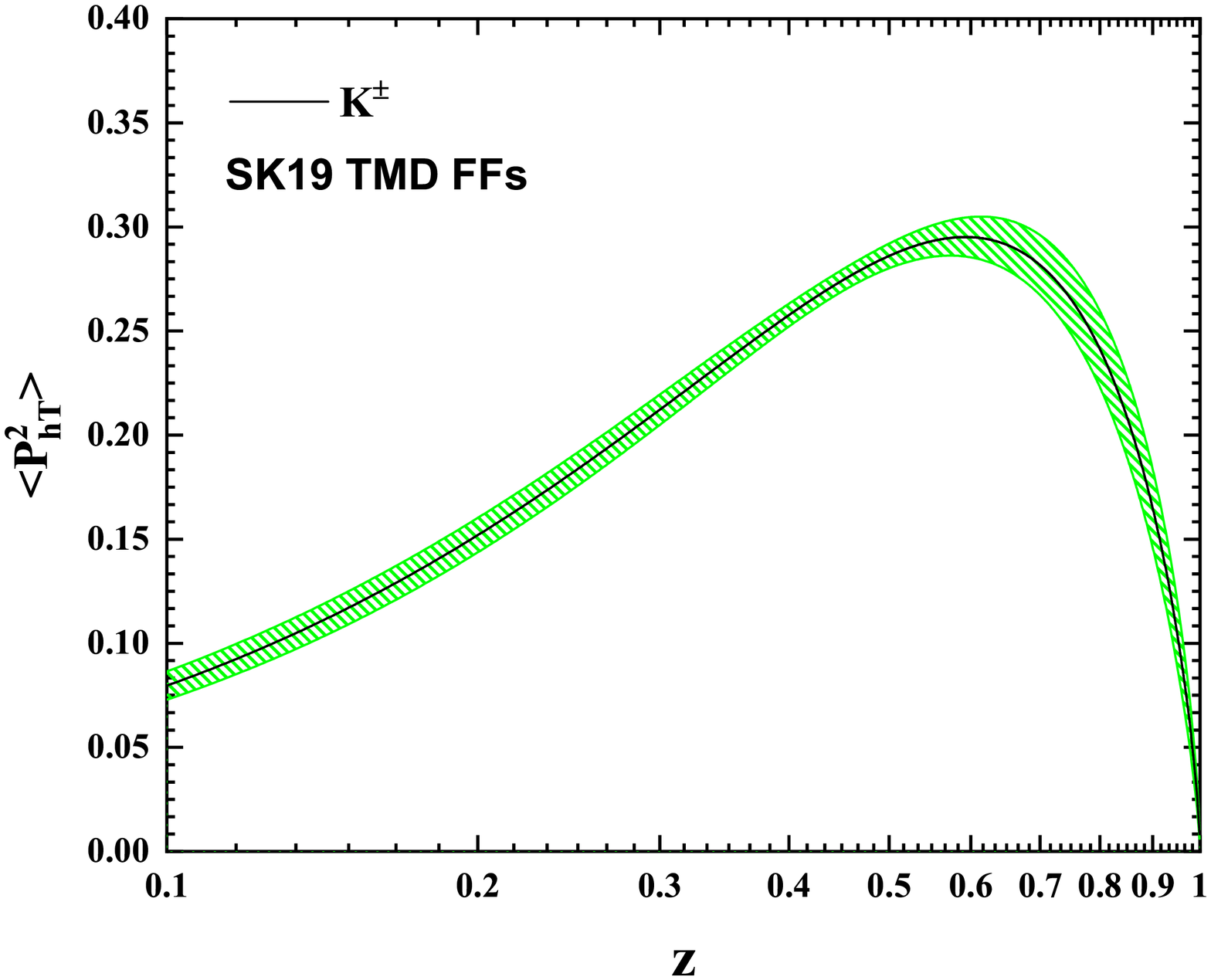}}  	
\resizebox{0.480\textwidth}{!}{\includegraphics{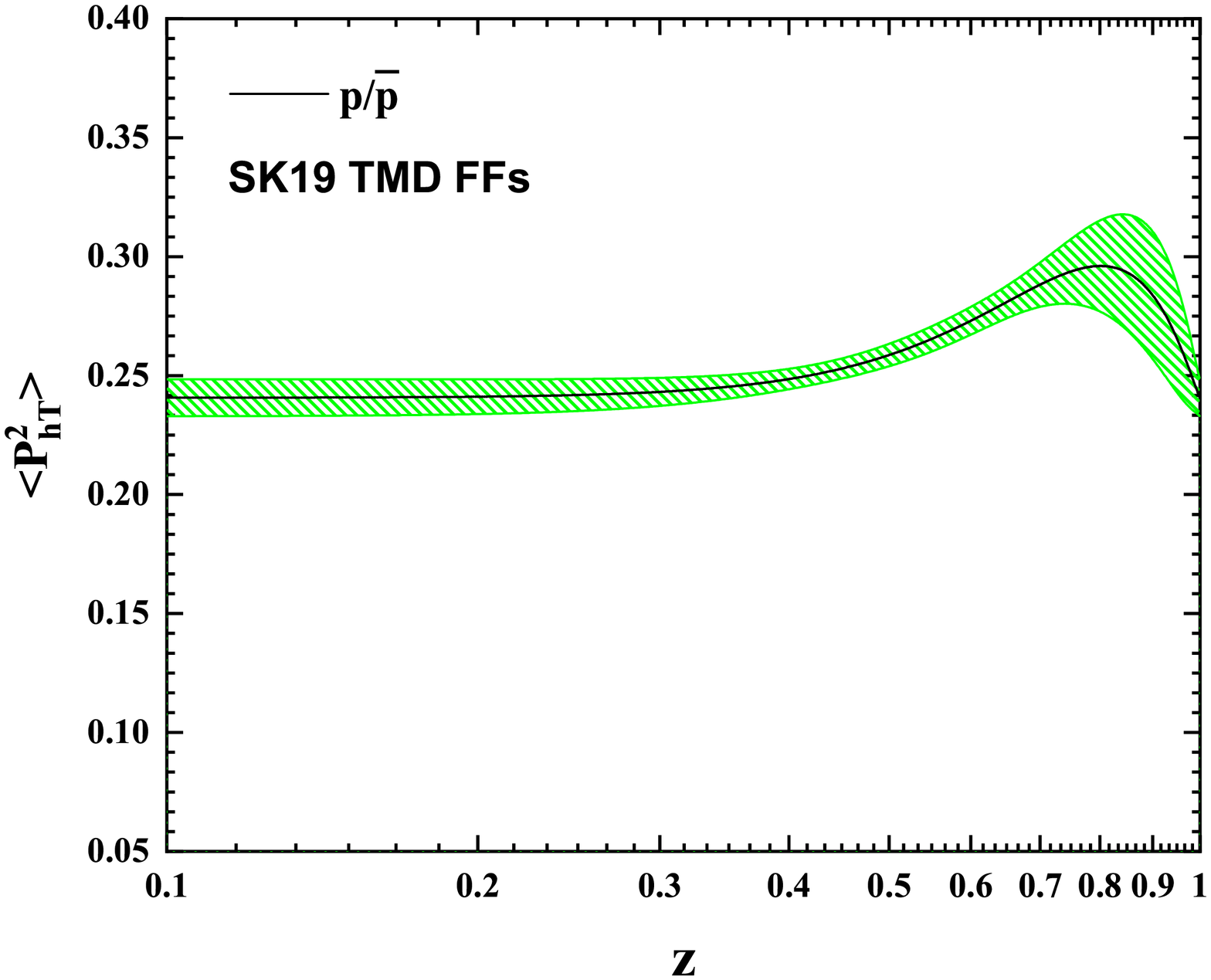}} 	
%\vspace{-6cm}
\begin{center}
\caption{{\small $\langle P_{hT}^2 \rangle$ distributions presented in Eq.~\eqref{eq:widthzdep} as a function of $z$. The plots shown are for the charged pion, Kaon and proton/antiproton FFs, respectively. The error bars correspond to the 1-$\sigma$ uncertainty at 68\% CL.  }}
\label{fig:TMD-FFs}
\end{center}
\end{figure*}

With the agreement between data and theory predictions established, we present now the results for the {\tt SK19 TMD FFs}  fits for the charged pion, kaon and proton/antiproton analyses along with their uncertainties. In order to present our results for the $\langle P_{hT}^2 \rangle$ distributions and discuss their shape for the different range of momentum fraction $z$, in Fig.~\ref{fig:TMD-FFs} we present $\langle P_{hT}^2 \rangle$ as a function of $z$. These plots are correspond to the charged pion, kaon and proton/antiproton TMD FFs, respectively with their 1-$\sigma$ uncertainty at $68\%$ CL. 

To conclude our discussions of the main properties of the {\tt SK19 TMD FFs} fits, we discuss in more details the shape and uncertainty bands of the extracted TMD FFs. In terms of central values, we can see that all distributions show a Gaussian shape which pick at $z \sim 0.6$ for $\pi ^\pm$ and $K^\pm$ and $z \sim 0.8$ for $p/\bar{p}$. The regions where the differences between these TMD FFs are the largest correspond to the small values of  $z$. The proton/antiproton TMD FFs shows a fixed pattern for the small values of $z$ and the charged kaon TMD FFs goes to zero at this region. Another differences between these distributions concerns the size of the TMD FFs uncertainty bands. We find that the proton/antiproton and charged kaon TMD FFs fits lead to a slight decrease in uncertainties while the charged pion  TMD  FF comes with a wider error bands, especially at small value of $z$; $z<0.3$. 

%------------------------------------------------
%------------------------------------------------

In order to illustrate the effects arising from the use of TMD datasets from Belle collaboration for $\pi^\pm$, $K^\pm$ and $p/\bar{p}$ hadrons in our analysis, in the last column of Table.~\ref{tab:datasets} presented in section~\ref{sec:exper}, we show the $\chi^2/{\rm dof}$ for each light hadrons. The value of $\chi^2/{\rm dof}$ clearly illustrate our fit quality for all hadrons individually. Considering the $\chi^2/{\rm dof}$ and at the level of individual light hadrons datasets, we find in most cases a good agreement between the experimental measurements from Belle experiment and the corresponding theory calculations. For the $p/\bar{p}$ TMD FFs, one can see a better fit quality than the $\pi^\pm$ and $K^\pm$. Moreover, we find that the fit quality is quite similar for the case of $\pi^\pm$ and $K^\pm$.

As a short summary, our results and the apparent fit quality shown by excellent $\chi^2$ values (see Table.~\ref{tab:datasets}), suggests that the {\tt SK19 TMD FFs }  QCD fits considering the Gaussian function can be used as a universal functions in different hadronization processes specially for SIDIS process.

%======================================================================
\section{Summary and Conclusions }\label{sec:conc}
%======================================================================

Very recently, Belle Collaboration at KEK has published the first measurements on $e^+ e^- \rightarrow hX$ differential cross sections in both $z$ and $P_{hT}$ space for charged pion, kaon and proton/antiproton~\cite{Seidl:2019jei}. Previously, there was no dataset on the transverse momentum dependence of the cross sections or multiplicities for extraction of the unpolarized TMD FFs for identified light hadrons. However, over the last few years, the measurements of the Collins asymmetries in $e^+e^-\rightarrow h_1 h_2X$ are performed by Belle and BABAR Collaborations, and hence, several dedicated analyses are used these datasets to calculate the polarized TMD FFs.  These very recent Belle datasets are the only available observables in SIA process which can be used,  for the first time, to determine the unpolarized TMD FFs for pin, kaon and proton from QCD fits. These new measurements could provide enough constrains on the energy fraction $z$ of the fragmentation process.

In this paper, we have presented {\tt SK19 TMD FFs}, the first determination of TMD FFs from a QCD analysis of very recent measurements of $e^+ e^- \rightarrow hX$ differential cross sections for charged pion, kaon and proton/antiproton by Belle Collaboration at KEK. With rapid improvements in the cross section measurements of SIA process, the focus of the QCD analysis should be shifted toward providing accurate determination of TMD FFs in the wide range of $z$ and $P_{hT}$. In the current study, according to a simple partonic picture, we assume that the cross section is factorized, and hence, the TMD FFs can be expressed considering the unpolarized collinear FF $D_{h/f}(z, Q^2)$ and a new term which depends on the $h^h(P_{hT})$. On the theory side, we have introduced a very flexible parameterization to better capture the variations in the $P_{hT}$ dependence of TMD FFs. We have assumed a Gaussian form for the TMD FFs. For the collinear FFs in our parameterization, we have used the most recent FFs of pion, kaon and proton/antiproton from {\tt NNFF1.0} Collaboration. 

A series of benchmark tests on kinematical cuts on the $z$ and $P_{hT}$ for various hadrons have been carried out, and the cuts resulted to the better fit agreement between the data and theory have been selected. We have shown that our Gaussian parametrization can successfully describe the data up to $P_{hT} \sim 0.9$ for charged pion, and $P_{hT} \sim 0.8$ for  charged kaon and $P_{hT} \sim 1$ for proton/antiproton. We examined the TMD FFs errors considering the ``Hessian'' approach.

As a final point, we should highlight again that, this research provides the first extensive extraction of TMD FFs of pion, kaon and proton/antiproton from QCD fit to the most recent differential cross sections measurements of $e^+ e^- \rightarrow hX$ from Belle Collaboration at KEK~\cite{Seidl:2019jei}. This first determination of unpolarized TMD FFs, reflects the importance and originality of this study and it is of great significance as it marks the first attempt to use the Belle measurements on $e^+ e^- \rightarrow hX$. 
In terms of directions for the future research, further work could be performed by considering the effect arising from the higher order correction. This analysis is restricted to the electron-positron annihilation processes, and hence, another possible area of future research would be to investigate the effect of another source of information  on the TMD FFs which mainly come from the SIDIS processes. In terms of future work, it would be interesting to repeat the analysis described here considering the mentioned improvements.  These are a number of important improvement which need to be taken into account and we plan to revisit our analysis in the near future. 

Parameterization for the sets of {\tt SK19 TMD FFs } presented in this work are available in the standard {\tt LHAPDF} format~\cite{Buckley:2014ana} from the author upon request.

%
%%%%%%%%%%%%%%%%%%%%%%%%%%%%%%%%%%%%%%%%%%%%%%%%%%%%%%%%%%%%%%%%%%%%%%%
\begin{acknowledgments}
%%%%%%%%%%%%%%%%%%%%%%%%%%%%%%%%%%%%%%%%%%%%%%%%%%%%%%%%%%%%%%%%%%%%%%%
%

Authors are thankful to Ralf Seidl, Ignazio Scimemi, Valerio Bertone and Mojtaba Mohammadi Najafabadi for many helpful discussions and comments.
Authors thank School of Particles and Accelerators, Institute for Research in Fundamental Sciences (IPM) for financial support of this project, and CERN theory department for their hospitality and support during the preparation of this paper. Hamzeh Khanpour also is thankful the University of Science and Technology of Mazandaran for financial support provided for this research.

\end{acknowledgments}
%

%\clearpage

\end{document}